\documentclass[aps, preprint, superscriptaddress, amsmath, amssymb, showpacs, showkeys]{revtex4-1}
\usepackage{graphicx}
\usepackage{dcolumn}
\usepackage{bm}
\usepackage{upgreek}
\usepackage{times}
\usepackage{mathrsfs}
\usepackage{multirow}
\usepackage{fancyhdr} 
\usepackage{hyperref} 
\usepackage{color}

\begin{document}

\title{ Configuration Dependence of Band Gap Narrowing and Localization in Dilute \texorpdfstring{GaAs$_{1-x}$Bi$_x$}{Ga(AsBi)} Alloys}

\author{Lars C. Bannow}\email{lars.bannow@physik.uni-marburg.de}%
\affiliation{%
 Department of Physics and Material Science Center, Philipps-Universit\"at, 35032 Marburg, Germany
}%

\author{Oleg Rubel}
\affiliation{%
 Department of Materials Science and Engineering, McMaster University, Hamilton, Ontario L8S 4L8, Canada
}%

\author{Phil Rosenow}%
\affiliation{%
  Department of Chemistry, Philipps-Universit\"at, 35032 Marburg, Germany
}%

\author{Stefan C. Badescu}
\affiliation{%
   Air Force Research Laboratory, Wright-Patterson AFB, Ohio 45433, USA
}%

\author{J{\"o}rg Hader}%
\affiliation{%
  NLCSTR Inc, 7040 N Montecatina Dr., Tucson, Arizona 85704, USA
}%

\author{Jerome V. Moloney}%
\affiliation{%
  NLCSTR Inc, 7040 N Montecatina Dr., Tucson, Arizona 85704, USA
}%

\author{Ralf Tonner}%
\affiliation{%
  Department of Chemistry, Philipps-Universit\"at, 35032 Marburg, Germany
}%

\author{Stephan W. Koch}%
\affiliation{%
 Department of Physics and Material Science Center, Philipps-Universit\"at, 35032 Marburg, Germany
}

\date{\today}

\begin{abstract}
Anion substitution with bismuth (Bi) in III-V semiconductors is an effective method for experimental engineering of the band gap $E_g$ at low Bi concentrations ($\leq 2\%$), in particular in gallium arsenide (GaAs). The inverse Bi-concentration dependence of $E_g$ has been found to be linear at low concentrations $x$ and dominated by a valence band-defect level anticrossing between As and Bi occupied $p$ levels. This dependence breaks down at high concentrations where empirical models accounting only for the As-Bi interaction are not applicable. Predictive models for the valence band hybridization require a first-principle understanding which can be obtained by density functional theory with the main challenges being the proper description of $E_g$ and the spin-orbit coupling. By using an efficient method to include these effects, it is shown here that at high concentrations $E_g$ is modified mainly by a Bi-Bi $p$ orbital interaction and by the large Bi atom-induced strain. This points to the role of different atomic configurations obtained by varying the experimental growth conditions in engineering arsenide band gaps, in particular for telecommunication laser technology.
\end{abstract}


\maketitle

%
%
\section{Introduction}\label{Sec:Introduction}

Alloying of gallium arsenide (GaAs) with bismuth Bi efficiently reduces the band gap $E_g$ and enhances the spin-orbit splitting~\cite{Oe_JJAP_37_1998,Francoeur_APL_82_2003,Broderick_SST_27_2012}, the magnitude of which exceeds the energy gap ($E_g$) of GaAs$_{1-x}$Bi$_x$ at the Bi content of $x_\text{Bi}\gtrsim9$\%~\cite{Usman_PRB_87_2013}. \textcite{Sweeney2011} suggested that these properties can lead to a suppression of non-radiative losses in GaAs$_{1-x}$Bi$_x$ alloys (bismides) by creating off-resonance conditions for the Auger recombination~\cite{Sweeney_JAP_113_2013}, thus making dilute bismide semiconductors a promising candidate for GaAs-based lasers in the telecommunication wavelength of 1.55~$\mu$m~\cite{Batool_chapter_GaAsBi_2013,Ludewig_JCG_370_2013}. Recent progress in device fabrication includes the demonstration of an electrically pumped laser with a Ga(AsBi) gain medium~\cite{Ludewig_APL_102_2013}, followed by continuing efforts to extend the emission to longer 
wavelengths relevant for telecommunication applications~\cite{Butkute_EL_50_2014}.

Previously, smaller reductions in GaAs bandgap were achieved in dilute nitrides, where nitrogen N was incorporated at the As sites~\cite{Gruening_PSSc_1_2004,Rubel_JAP_98_2005,Jandieri_PRB_86_2012,Karcher_JL_133_2013}. In that case the $E_g$ reduction was understood as the hybridization (anticrossing) of unoccupied nitrogen $s$ orbitals with the host conduction band~\cite{Virkkala_PRB_85_2012,Virkkala_PRB_88_2013b}, giving localized states responsible for the conduction band tail detected experimentally. Two important distinctions occur between bismides and nitrides: in bismides the reduction of $E_g$ is explained by the hybridization of the {\it valence band} with $\it occupied$ Bi $p$ orbitals, and at high Bi concentrations compositional disorder plays a major role. On the one hand, the valence band hybridization in dilute bismides was inferred from transport measurements which showed a reduction of hole mobility by an order of magnitude compared to the host GaAs~\cite{Kini_PRB_83_2011,
Nargelas_APL_98_2011,Beaton_JAP_108_2010}, while the electron mobility is much less affected~\cite{Kini_PRB_83_2011,Cooke_APL_89_2006}. The hybridization mechanism is supported also by numerous electronic structure calculations performed at different levels: linear combination of atomic orbitals~\cite{Usman_PRB_84_2011,Usman_PRB_87_2013}, density functional theory (DFT)~\cite{Deng_PRB_82_2010,Virkkala_PRB_88_2013} and unfolding of DFT band structure~\cite{Rubel_PRB_90_2014}.
On the other hand, both spatial and valence band tail disorder from Bi incorporation had to be invoked to interpret photoluminescence (PL) experiments. The latter gave a broad low-temperature line width and a non-monotonous temperature dependence of both the PL peak position and the PL line width~\cite{Imhof_APL_96_2010,Imhof_PSSB_248_2010,Imhof_APL_98_2011,Shakfa_JAP_114_2013,Shakfa_JAP_025709_2015}, and had to be explained by two-scale disorder models.

Furthermore, experimental evidence for Bi cluster formation was reported~\cite{Imhof_APL_96_2010, Puustinen_JAP_114_2013, Ciatto_PRB_78_2008}. Clusters are observed in samples grown at low temperatures while they are absent in samples grown at higher temperatures~\cite{Puustinen_JAP_114_2013}. The occurrence of clusters depends on the Bi concentration.~\cite{Ciatto_PRB_78_2008} For concentrations $x\leq1.2~\%$ the Bi atoms seem to be randomly distributed and interactions between to have a reduced role, whereas Bi pair and cluster formation seem to occur at concentrations $x\ge1.9~\%$, reducing the average Ga-Bi bond length~\cite{Ciatto_PRB_78_2008,Usman_PRB_84_2011} and giving rise to a Bi defect level in the band gap~\cite{Usman_PRB_84_2011}. The relation between Bi-Bi interactions and localized defect levels has been shown in simulations of selected next-nearest neighbor configurations~\cite{Virkkala_PRB_88_2013}. The relation between Bi-Bi interactions in pair, triplet, and 
other random configurations at high concentrations on the one hand, and the valence tail level disorder on the other hand, does not have yet a microscopic understanding.

In this work, we use accurate DFT first principle calculations to investigate not only the $E_g$ dependence on Bi concentration, but also the mechanism of valence band tail formation and its relation to various Bi complexes. Previous DFT models considered either single Bi atoms or two neighboring Bi atoms distributed periodically by means of simulation supercells. Here we demonstrate the strong dependence of the $E_g$ renormalization on the Bi cluster structure for a given concentration and provide an insight from charge accumulation at the Bi sites. Then we relate this charge localization to the Bi-Bi $p$ {\it wavefunction} overlap, which turns out to give a strong dependence of localized level energies on the distance between the two Bi atoms. Finally, we use a band unfolding technique~\cite{Rubel_PRB_90_2014} to show how these energy distributions determine the band tails and disorder observed in PL experiments.

In the following Section~\ref{Sec:Method}, we describe the calibration and validation of our DFT methods. Section~\ref{Sec:Results} provides the results and goes into the details of the points mentioned above.


%
%
\section{Computational details}\label{Sec:Method}
It has been shown that the accurate description of the energy band gap $E_g$ and of the valence band spin-orbit split-off in III-V semiconductors by DFT is sensitive to the choice of the density functional or correction potentials for describing screening effects and to the incorporation of spin-orbit coupling (SOC)~\cite{Kim_PRB_80_2009,Kim_PRB_82_2010}. We include these effects in two sets of calculations that validate each other: one with the Projector Augmented-Wave (PAW) pseudopotential method~\cite{Bloechl_PRB_50_1994,Kresse_PRB_59_1999} implemented in the Vienna Ab-initio Simulation Package~\cite{Kresse_PRB_47_1993,Kresse_PRB_49_1994,Kresse_CMS_6_1996,Kresse_PRB_54_1996} (VASP), and one with the all-electron Linearly Augmented Plane Waves (LAPW) method implemented in the WIEN2k package~\cite{Blaha_2001}. The former is computationally more efficient for large systems, whereas the latter is very accurate but efficient for small systems. This standard comparison between the two DFT packages 
has been done 
before for binary III-V compounds~\cite{Kim_PRB_82_2010} and yielded a good agreement between WIEN2k and VASP. We find below that this holds also for the alloys considered here.

The SOC and the accurate description of $E_g$ increase the computational cost and limits the modeling of defects and alloys to a few-hundred atom supercells. The method of choice for such systems uses PAW, made efficient by replacing the rapidly oscillating portion of the valence electron wavefunctions close to the atomic cores by smooth functions. An additional increase in efficiency of the PAW method is obtained from atomic pseudopotentials (PP), which replace the deep localized atomic levels with atomic cores. It was demonstrated that the PAW-PP method implemented in VASP describes accurately the III-V semiconductor band structures~\cite{Kim_PRB_82_2010} and it has been used for dilute nitrides~\cite{Virkkala_PRB_85_2012,Virkkala_PRB_88_2013b} and bismides~\cite{Virkkala_PRB_88_2013}. We used the PAW-PPs in conjunction with SOC and with the Tran-Blaha Modified Becke Johnson potential (TBmBJ) implemented in VASP for the majority of the calculations in this paper.

In addition to the VASP calculations, we performed a series of calculations with the all-electron WIEN2K DFT package, which uses the LAPW method applied to localized basis sets. This was done mainly to cross-check and validate the accuracy of the PAW-PP calculations, and also to demonstrate the effect of Bi on forming alloy effective bands and band tails. For the latter we used the package \texttt{fold2Bloch}~\cite{Rubel_PRB_90_2014} based on WIEN2K output.  We show the comparison between the two methods in Tab.~\ref{Table:E} and Tab.~\ref{Table:B} below. First, we describe the tuning of both methods for the systems at hand.

\subsection{PAW-Pseudopotential Calculations}\label{Sec:Method:Pseudopotentials}
The first step in the PAW-pseudopotential calculations was to determine the lattice constant of the GaAs primitive cell and to choose the density functional that provides an $E_g$ closest to experiment. For that, we compared the local density approximation (LDA)~\cite{Perdew_PRB_45_1992} to the generalized gradient approximation (GGA) in the Perdew, Burke and Ernzerhof parametrization (PBE)~\cite{Perdew_PRL_77_1996,Perdew_PRL_78_1997} and to the modified PBE functional GGA-PBEsol~\cite{Perdew_PRL_100_2008}. For the latter two functionals we included the van-der-Waals D3 correction method with Becke-Johnson damping~\cite{Grimme_JOCP_132_2010,Grimme_JOCC_32_2011}. We used a $8\times8\times8$ k-mesh Brillouin zone sampling~\textcite{Monkhorst_PRB_13_1976} and a plane wave cutoff energy of $510\,$eV. In the next step, the structures obtained were used as inputs for the band structure calculations. In the latter we applied the bandstructure correction implemented in the efficient Tran-Blaha 
modified 
Becke-Johnson potential (TBmBJ)~\cite{Tran_PRL_102_2009}. The results summarized in Tab.~\ref{Table:E} show that the GGA-PBE$+$TBmBJ combination provides parameters closest to experiment, therefore we used it for the supercell calculations.

   In the third step, we used the resulting value of $a_0 = 5.668$~{\AA} to construct supercells of sizes $2\times2\times2$, $3\times3\times3$, and $4\times4\times4$, containing $16$, $54$ and $128$ atoms, respectively. In each case the k-point mesh grid was scaled down accordingly to $4\times4\times4$, $3\times3\times3$ and $2\times2\times2$. The supercell sizes were frozen at their multiples of $a_0$ irrespective of the chemical composition. Atomic positions in Bi-containing alloys were relaxed internally by minimizing Hellmann-Feynman forces acting on atoms below 20~meV/{\AA}.
\begin{table}
    \caption{Tuning of VASP PAW-PP calculations: equilibrium lattice constant $a_0$, band gap $E_\text{g}$, spin-orbit splitting $E_\text{SO}$ and the energy offset between valleys in the conduction band of GaAs obtained from PAW-DFT calculations using LDA, GGA-PBE and GGA-PBEsol potentials. The lattice constant was calculated with van-der-Waals D3 correction in the case of GGA-PBE and GGA-PBEsol potentials. The band structure parameters were calculated by adding the TBmBJ potential~\cite{Tran_PRL_102_2009}. Results are also compared to the corresponding experimental values extrapolated to zero temperature.}\label{Table:E}
    \begin{ruledtabular}
        \begin{tabular}{l c c c c}
            Parameter & LDA & GGA-PBE & GGA-PBEsol & Expt. (0~K) \\
            \hline
            $a_0$ [\AA]                     & 5.606 & 5.668 & 5.592 & 5.648~\cite{Madelung2004}\\
            $E_\text{g}$ (eV)                & 1.77 & 1.44 & 1.74 & 1.52 ~\cite{Blakemore_JAP_53_1982} \\
            $\text{L}_{6c}-\Gamma_{6c}$ (eV) & 0.04 & 0.25 & 0.06 & 0.30 ~\cite{Blakemore_JAP_53_1982}\\
            $\text{X}_{6c}-\Gamma_{6c}$ (eV) & 0.15 & 0.59 & 0.19 & 0.46 ~\cite{Blakemore_JAP_53_1982}\\
            $E_\text{SO}$ (eV)               & 0.32 & 0.31 & 0.31 & 0.33 ~\cite{Lautenschlager_PRB_35_1987}\\
        \end{tabular}
    \end{ruledtabular}
\end{table}

\subsection{All-electron calculations}\label{Sec:Method:All-electron}

Similarly to the PAW-PP method tuning above, we tuned the all-electron calculations performed with WIEN2k by finding the combination of functionals and correcting potentials that describes best the lattice parameter and the energy gap $E_g$. Besides GaAs properties, an additional validation for the WIEN2k calculations was to find the bandstructure of GaBi and compare it against the state-of the art model available in the literature.

The muffin tin radii $R^\text{MT}$ where set to $2.17$,~$2.06$, and $2.28$~bohr for Ga, As and Bi, respectively. The product $R^\text{MT}_\text{min}K_\text{max}=7$, which determines the accuracy of a plane wave expansion of the wave function, was used throughout the calculations. For single unit-cell calculations the Brillouin Zone was sampled using $8\times8\times8$ mesh. The atomic positions were optimized by minimizing Hellmann-Feynman forces acting on atoms below $2$~mRy/Bohr. The choice of exchange correlation functional was based on preliminary study of the band structure of GaAs. The lattice constant and the band structure were calculated self-consistently using the~\textcite{Wu_PRB_73_2006} (GGA-WC) and the~\textcite*{Perdew_PRL_77_1996} (GGA-PBE) versions of the GGA, as well as the LDA~\cite{Perdew_PRB_45_1992}. The TBmBJ potential~\cite{Tran_PRL_102_2009} was applied in order to improve accuracy for the band gaps. The results are summarized in Tab.~\ref{Table:B}. The band gap of 1.62~eV was 
obtained 
with LDA-TBmBJ for GaAs at the \textit{experimental} geometry~\cite{Tran_PRL_102_2009}. Tab.~\ref{Table:B} shows that for the all-electron calculations, the combination of GGA-WC with TBmBJ provides the best description for the uppermost part of the valence band and for the lowest sets of conduction band minima in GaAs. Therefore, we used this combination for the band structure calculations of GaAs$_{1-x}$Bi$_x$ alloys performed with WIEN2k. It should be noted that the poor performance of LDA and GGA-PBE can be partly attributed to the error in the lattice constant, which is discussed in detail by~\textcite{Haas_PRB_79_2009}.

We built the supercells as multiples of the two-atom primitive cell basis instead of the conventional eight-atom crystallographic cell, as required for calculating the effective band structure of an alloy. The GGA-WC self-consistent lattice constant of $a_0=5.660$~{\AA} from above was used for the host GaAs. The Brillouin Zone sampling was downscaled to $2\times2\times2$ for a $128$-atom supercell used in the example of effective-bandstructure Bloch spectral weight shown later.
The comparison of tables Tab.~\ref{Table:E} and Tab.~\ref{Table:B} shows that the GGA approximation with the TBmBJ correction gives similar results for $a_0$ and $E_g$ in VASP and WIEN2K, albeit the former has to use the PBE parametrization, while the latter the WC version.

\begin{table}
    \caption{Equilibrium lattice constant $a_0$, band gap $E_\text{g}$, spin-orbit splitting $E_\text{SO}$ and the energy offset between valleys in the conduction band of GaAs obtained from self-consistent all-electron DFT calculations using various exchange correlation functionals. The band structure parameters were calculated by adding the TBmBJ potential~\cite{Tran_PRL_102_2009}. Results are also compared to the corresponding experimental values extrapolated to zero temperature.}\label{Table:B}
    \begin{ruledtabular}
        \begin{tabular}{l c c c c c}
            Parameter & GGA-WC & GGA-PBE & LDA & Expt. (0~K) \\
            \hline
            $a_0$ (\AA)                & 5.660 & 5.737 & 5.609 & 5.648~\cite{Madelung2004} \\
            $E_\text{g}$ (eV)          &  1.53 &  1.22 &  1.73 &  1.52 ~\cite{Blakemore_JAP_53_1982} \\
            $\text{L}_{6c}-\Gamma_{6c}$ (eV) &  0.18 &  0.38 &  0.05 &  0.30 ~\cite{Blakemore_JAP_53_1982}\\
            $\text{X}_{6c}-\Gamma_{6c}$ (eV) &  0.48 &  0.87 &  0.21 &  0.46 ~\cite{Blakemore_JAP_53_1982}\\
            $E_\text{SO}$ (eV)         &  0.29 &  0.29 &  0.30 &  0.33 ~\cite{Lautenschlager_PRB_35_1987}\\
        \end{tabular}
    \end{ruledtabular}
\end{table}

\begin{figure}
	\includegraphics[width=0.3\textwidth]{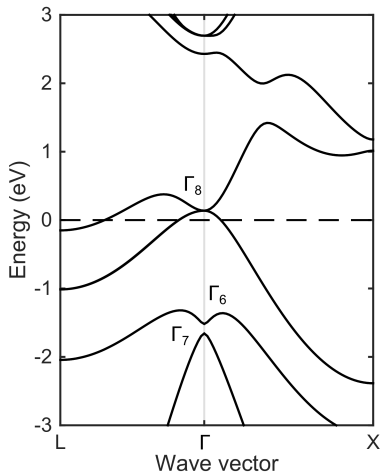}\\
	\caption{Band structure of zinc-blend GaBi obtained with GGA-WC-TBmBJ. Energies are plotted relative to the Fermi energy.
\label{Fig:C}}
\end{figure}

A further validation of our calculations is to show that the bandstructure of GaBi can be predicted accurately.
The information on the band structure of GaBi is scarce.
To date, calculations by~\textcite{Janotti_PRB_65_2002} performed with LDA+$C$ are considered state of the art. LDA+$C$ is an empirical correction in the form of an additional atom-dependent radial potential, which is introduced in order to overcome shortcomings of LDA band structure~\cite{Christensen_PRB_30_1984,Wei_PRB_57_1998}. The potential parameters for LDA+$C$ are selected based on experimental band gaps and lattice constants of binary compounds, which is problematic in the case of GaBi. Therefore it will be useful to present results of all-electron DFT calculations for the band structure of GaBi obtained with TBmBJ correction.

To model the GaBi bandstructure, we used the GGA-WC exchange correlation functional as explained above. In the first step, the lattice constant of zinc-blend GaBi is optimized taking into account spin-orbit coupling. The result found here of $a_0=6.368$~{\AA} is consistent with previous DFT calculations: 6.324, 6.28 and 6.47~{\AA}~\cite{Janotti_PRB_65_2002,Ferhat_PRB_73_2006}. This gives a lattice mismatch of 12\% relative to GaAs (Tabs.~\ref{Table:B} and \ref{Table:A}), which hints that Bi atoms embedded in GaAs host lattices will give rise to large strains. This will be confirmed in the next section. Here, we used this $a_0$ to calculate the GaBi band structure shown in Fig.~\ref{Fig:C}. Relativistic effects play an important role in the electronic structure of GaBi that is evident from a large spin-orbit splitting of $E_\text{SO}=1.80$~eV. We find an inverted band structure with the $\Gamma_{6c}$ state positioned energetically below the $\Gamma_{8v}$ state, in agreement with~\textcite{Janotti_PRB_65_2002}. This arrangement results in a negative band gap of $E_\text{g}=-1.65$~eV at $\Gamma$ point, which is comparable to the LDA+$C$ calculations that yield the energy gap of $-1.45$~eV~\cite{Janotti_PRB_65_2002}. This additional test validates our WIEN2K calculations for Bi-containing compounds.

%
%
\section{Results and discussion}\label{Sec:Results}

\subsection{Comparison of strain and chemical effects in bandgap bowing}\label{SSec:Bi-distort}

The 12\% lattice mismatch between GaAs and GaBi binary compounds described above is consistent with the large difference between the covalent radii of As and Bi (1.19 \textit{vs}. 1.48~{\AA},~\cite{Cordero_DT__2008}). Using our computational methods we find that a Bi atom in $4\times4\times4$ GaAs host supercell gives a relaxed Ga-Bi bond longer by 7.5\% than the Ga-As bond in pristine GaAs. This is the second largest magnitude of the local distortion field introduced in GaAs after nitrogen (Tab.~\ref{Table:A}). The valence band of GaAs is dominated by a deep As $s$ level and by three equal-energy As $p$ levels which give rise to the top of the valence band via overlaps between primitive cells~\cite{Gray_NatureMat_2011}. This overlap and therefore the bandstructure can be perturbed by lattice strain as described by band deformation potentials~\cite{Nolte_PRL_4_1987}. Including a Bi atom imposes such a strain in the host lattice, which is partially responsible for the bandgap variations in 
bismides. We call this the {\it strain effect}. An additional perturbation is due to the higher energy of the $p$ valence orbitals of Bi by comparison to the $p$ As orbitals, which we call the {\it chemical effect}.
\begin{table}
    \caption{Strain of anion-cation bond lengths $(r)$ in the nearest-neighbour shell of isovalent group-V impurities in GaAs.}\label{Table:A}
    \begin{ruledtabular}
        \begin{tabular}{l c}
            Compound\footnote{Results for N, P and Sb are adopted from Ref.~~\citenum{Rubel_PRB_78_2008}.}  & $\epsilon = r/r_0 - 1$\footnote{The strain is calculated with respect to the equilibrium bond length $(r_0)$ in GaAs.} \\
            \hline
            GaAs:N & $-0.155$\\
            GaAs:P & $-0.025$\\
            GaAs:Sb & $+0.053$\\
            GaAs:Bi & $+0.075$\\
        \end{tabular}
    \end{ruledtabular}
\end{table}

To delineate the strain and the chemical effects on the $E_g$ bowing in bismides we compare calculations  that take into account the Bi chemistry with models where the Bi atoms are replaced back with As anions. We analyze two cases: first, periodic arrangements obtained with single Bi atoms in supercells of increasing sizes $2\times2\times2$, $3\times3\times3$ and $4\times4\times4$; and second, a random distribution of Bi atoms in a supercell of size $4\times4\times4$. In all cases, three electronic structures were obtained with the following models: ({\it i.})~{\it frozen lattice}, with atomic positions frozen to the host lattice, disregarding the local lattice distortions due to Bi; ({\it ii.})~{\it relaxed lattice}, with atomic positions relaxed by minimizing the forces arising to Bi incorporation, keeping the supercell size fixed; and ({\it iii.})~{\it distorted pristine lattice}, where the positions of atoms are taken from {\it ii.} and Bi is replaced back by As. The latter case allows to isolate 
changes in the host band structure caused solely by the lattice distortions~\cite{Kent_PRB_64_2001}.

\begin{figure}
	\includegraphics[width=0.36\textwidth]{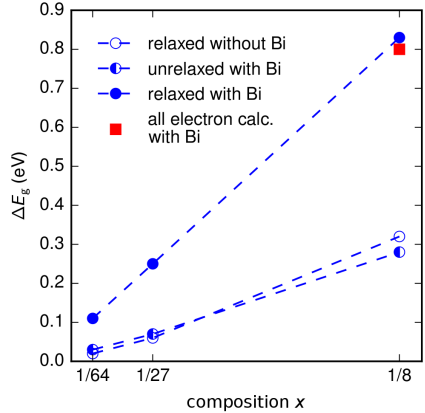}\\
	\caption{Variation of the band gap where $\Delta E_\text{g} = E_\text{g}^\text{GaAs}-E_\text{g}^\text{Ga(AsBi)}$ in GaAs$_{1-x}$Bi$_x$ as a function of composition $x$. Local atomic displacements induced in GaAs host lattice by Bi are a significant factor that contributes to the band gap bowing. To demonstrate the good agreement between all-electron calculations and pseudopotential calculations, we included a data point from an all-electron calculation (red square, Sec.~\ref{Sec:Method:All-electron}). The all-electron bandstructure calculation was based on structure files obtained by the PAW-PP method. All other results were obtained entirely with the PAW-PP method (Sec.~\ref{Sec:Method:Pseudopotentials}).}\label{Fig:B}
\end{figure}

The calculations with a single Bi atom per supercell correspond to idealized crystals with unit cells of $16$, $54$, and $128$ atoms, respectively. The resulting $E_g$ bowing is given as a function of composition $x$ in Fig.~\ref{Fig:B}. These were obtained with the PAW method (Sec.~\ref{Sec:Method:Pseudopotentials}), and we added a data point from an all-electron calculation (Sec.~\ref{Sec:Method:All-electron}) to show that the results are in good agreement. While at large concentrations $x\approx 0.125$ the {\it frozen lattice} model
shows a sizable $E_g$ bowing, but less than half of that from
the {\it relaxed lattice} model, at low concentrations it shows a much smaller fraction. The {\it distorted pristine lattice} model gives $E_g$ bowing very similar to the {\it frozen lattice} model for all concentrations. The summation of the two is less than the $E_g$ bowing observed in the {\it relaxed lattices}, with a pronounced difference at small $x$.

\begin{figure*}
	\includegraphics[width=1.0\textwidth]{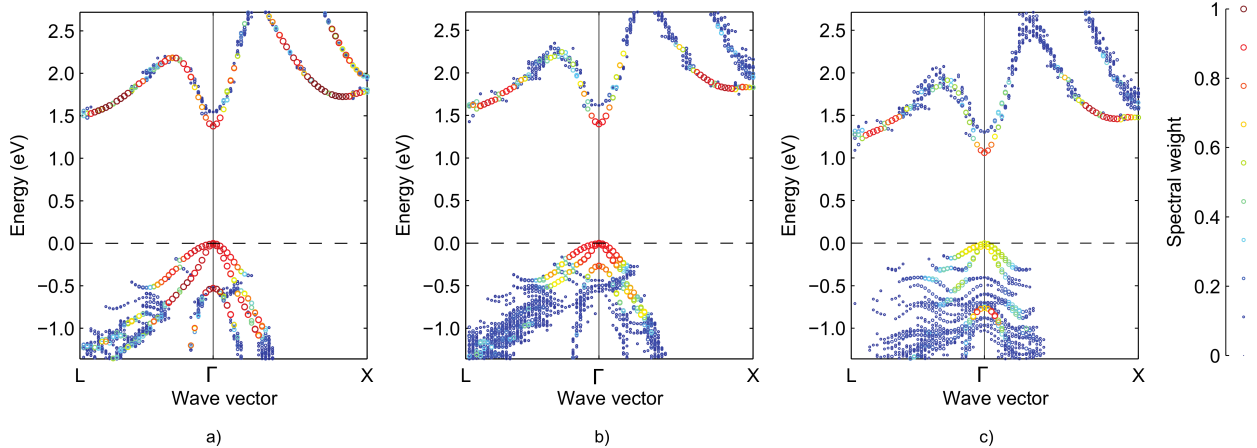}\\
	\caption{Effective band structure of a random Ga$_{64}$As$_{58}$Bi$_{6}$ supercell unfolded to a primitive Bloch representation. Disorder effects are  partitioned into the {\it chemical effect} (a) and the {\it strain effect} due to the size mismatch between Bi and As atoms (b). Panel (c) represents the total effect, which is larger than the sum of the two. Details for the separation of the effects are given at the beginning of  Sec.~\ref{SSec:Bi-distort}. The energy reference is taken at the valence band edge. The Bloch spectral weight is represented by colour and the symbol size. Points with the spectral weight of less than 5\% are filtered out.}\label{Fig:A}
\end{figure*}

The results for a random distribution of Bi atoms in a supercell with $128$ atoms are shown in Fig.~\ref{Fig:A}.
These correspond to the composition of $9.4$\% Bi and were obtained with the all-electron method above (Sec.~\ref{Sec:Method:All-electron}). This large composition is relevant for telecom lasers with the emission wavelength of $1.55$~$\mu$m and is close to the crossover between the band gap and spin-orbit splitting that takes place in GaAs$_{1-x}$Bi$_x$ at $x\approx9-10$\%~\cite{Usman_PRB_84_2011,Usman_PRB_87_2013}. The alloy was represented by six As atoms randomly substituted by Bi. Three sets of calculations {\it i}-{\it iii} are performed according to the description above. The {\it frozen lattice} model applied to this alloy [Fig.~\ref{Fig:A}~(a)] gives a mild perturbation of the pristine GaAs band structure induced by Bi disorder. The most noticeable changes occur in the valence band, such as an enhanced spin-orbit splitting and smeared Bloch character of states located well below the Fermi energy. The Bloch character of the conduction band remains almost unaffected. The {\it distorted pristine lattice}
 model [Fig.~\ref{Fig:A}~(b)] shows a disorder in the conduction band from local lattice distortions but, more importantly, it shows that the valence band is perturbed to the extent that it looses its Bloch character for states with the energy $E<-0.5$~eV, where it becomes hard to distinguish between heavy and light holes. Nevertheless, in this model the uppermost valence band preserves its Bloch character. The combined strain and chemical effects are seen in the {\it relaxed lattice model} [Fig.~\ref{Fig:A}~(c)], which displays profound changes in the valence band: the Bloch character is deteriorated down to $60\%$ even for the edge of the valence band at the $\Gamma$ point. Such a low value is indicative of localization effects in the valence band, and it is correlated with an $E_g$ bowing larger than the sum of those in Figs.~\ref{Fig:A}~(a,b). This corroboration of strain and chemical effects in decreasing $E_g$ is consistent with the previous observation from Fig.~\ref{Fig:B}, and the change in $E_g$ 
for the alloy is quantitatively consistent with those from PAW method. We also observe in Fig.~\ref{Fig:A}~(c) that the heavy hole band becomes progressively weaker (smaller spectral weight) and eventually disappears while moving deeper into the valence band. In spite of the disorder, the split-off band can be clearly identified. The conduction band is overall less disturbed and retains 80\% of its $\Gamma$ character.

The trends noticed here for a random distribution of Bi atoms are consistent with the results obtained using a tight binding model~\cite{Usman_PRB_87_2013,Usman_PRB_84_2011}. In addition, the observations made for both the random and the regular distributions point towards a hybridization of Bi $p$ orbitals mainly with the heavy and light hole bands throughout a large energy interval including the valence band edge, and also to interactions between Bi atoms. In the next paragraph we provide an understanding of both effects by looking at clusters of Bi atoms.

\subsection{Bi complexes}\label{Sec:Results:Clustering}
The random distribution in Fig.~\ref{Fig:A} for six Bi atoms among the $64$ sites of a $4\times4\times4$ supercell corresponds to a concentration of $x=9.37\%$, between the ordered-Bi arrangements with $x=3.7\%$ and $12.5\%$ in Fig.~\ref{Fig:B}. The latter two imply Bi-Bi distances of three, respectively two lattice spacings in all directions, whereas random distributions like that in Fig.~\ref{Fig:A} can include nearest neighbors, next-nearest neighbors, etc. A given concentration can be modeled with more than one Bi atom per supercell, e.g., $x=12.5\%$ can be modeled as above with one Bi atom per $2\times2\times2$ supercell, or with eight Bi atoms in a $4\times4\times4$ supercell, etc., which is expected to produce a distribution of data points in Fig.~\ref{Fig:B}. This is exactly what we find in Fig.~\ref{Fig:E} below, obtained with the approach described in the next paragraph. The following models have been analyzed in order to understand some aspects of Bi-atom clustering and to make 
initial steps towards interpreting experiments, in particular for high concentrations $x$.  We address the $E_g$ bowing for combinations of two, three, and four Bi atoms, respectively, and then we analyze the electronic bandstructure of two Bi atoms in several relative positions. For these calculations we used the PAW method (Sec.~\ref{Sec:Method:Pseudopotentials}) and included the atomic relaxation inside the supercell. 

We considered several different Bi complexes in a $128$-atom supercell and observed the $E_g$ bowing. We took the first Bi atom to be at the origin and specified the other Bi atoms by their positions relative to it through $(m_1,m_2,m_3)$=$m_1\vec{a}_1+m_2\vec{a}_2+m_3\vec{a}_3$. Here, $\vec{a}_1$, $\vec{a}_2$ are $\vec{a}_3$ are the two-atom primitive lattice vectors. First, we constructed three arrangements for each concentration $x=3.13$,~$4.69$,~$6.25\%$, shown in Tab.~\ref{Table:C}: chains along axis $[100]$, chains along axis $[111]$ and clusters (in which all Bi are closest to origin along $[100]$ and $[111]$ directions).
In addition, for this $128$-atom supercell we used the ATAT package~\cite{Walle_Calphad_26_2002,Walle_Calphad_33_2009} to obtain \textit{special quasirandom structures} (SQS)~\cite{Zunger_PRL_65_1990,Walle_Calphad_42_2013}. These structures are as close energetically as possible with periodic supercells to the true disordered state. We chose the pair length to include third nearest neighbours, the triplet length to include second nearest neighbours and the quadruplet length to include nearest neighbours. The difference from the correlation functions of the supercells we obtained to the correlation functions of the true disordered state are all smaller than $0.025$.


\begin{table}
    \caption{Effect of different Bi arrangements in a 128-atom supercell on the band gap $E_\text{g}$~(eV) for two, three and four Bi atoms. $\Delta E_\text{TOT}$~(eV) is the total energy difference of the supercells with respect to total energy of the SQS for each amount of Bi atoms.
    }\label{Table:C}
    \begin{ruledtabular}
        \begin{tabular}{c c c c c c}
            &Arrangement & 2 atoms & 3 atoms & 4 atoms & $\Delta E_g/x$~[meV/$\%$Bi] \\
            \hline
            &$[100]$ chain & \{(0,0,0), & \{(0,0,0), & \{(0,0,0),& $144\pm 23$\\
            &            & (2,0,0)\} & (1,0,0), & (1,0,0),& \\
            &            &         & (2,0,0)\} & (2,0,0),& \\
            &            &         &         & (3,0,0)\}& \\
$E_\text{g}$&            & 1.15    & 0.87    & 0.52   & \\
$\Delta E_\text{TOT}$&   & 0.39    & 0.27    & 0.67   & \\
            \hline
            &$[111]$ chain & \{(0,0,0), & \{(0,0,0), & \{(0,0,0),& $41\pm5$\\
            &            & (2,2,2)\} & (2,2,2), & (1,1,1),& \\
            &            &         & (3,3,3)\} & (2,2,2),& \\
            &            &         &         & (3,3,3)\}& \\
$E_\text{g}$&            &  1.28   & 1.23    &  1.19  & \\
$\Delta E_\text{TOT}$&   & 0.31    & 0.18    & 0.33   & \\
            \hline
            &clustered& \{(0,0,0),& \{(0,0,0), &  \{(0,0,0),&$90\pm10$\\
            &        & (1,0,0)\} & (1,0,0), &  (1,0,0),&\\
            &        &  (pair) & (0,1,0)\} &  (0,1,0),&\\
            &        &         & (trimer)&  (0,0,1)\}&\\
            &        &         &         &(tetramer)&\\
$E_\text{g}$&        &  1.12   & 0.96    & 0.90    & \\
$\Delta E_\text{TOT}$&& 0.36   & 0.25    & 0.73    & \\
             \hline
            &SQS    &          &         &         & $45\pm 7$\\
$E_\text{g}$&        & 1.32    & 1.19    & 1.18    & \\
$\Delta E_\text{TOT}$&&0.00    & 0.00    & 0.00    & \\
        \end{tabular}
    \end{ruledtabular}
\end{table}


\begin{figure}
	\includegraphics[scale=.7]{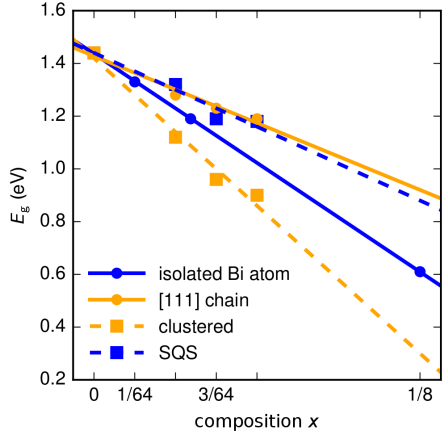}\\
	\caption{Variation of the band gap in GaAs$_{1-x}$Bi$_x$ as a function of composition for different types of arrangements. Included are the isolated atom arrangement from Fig.~\ref{Fig:B} (blue dots), SQS data from 128 atom supercells (blue squares), the $[111]$ chain data from Tab.~\ref{Table:C} (orange dots) and the cluster (pair, triplet, tetramer) data from Tab.~\ref{Table:C} (orange squares). Straight lines were fitted through the data points and the value for pure GaAs.}\label{Fig:E}
\end{figure}

Tab.~\ref{Table:C} shows the obtained band gaps $E_g$ for all arrangements considered, along with the total energy difference $\Delta E_\text{TOT}$ between each arrangement and the SQS structure at a given concentration. In all cases, the SQS are energetically most stable. The other arrangements for a given $x$ differ only slightly in their total energy (less than $0.1\,$eV), with the $[111]$ chain preferred after the SQS. The exception is the $[111]$ chain arrangement of four atoms, which is considerably more stable than the chain and tetramer for this concentration. The last column in the table shows the slope $\Delta E_g/x$ of the band gap bowing for each of these arrangement types, obtained from the linear interpolation shown in Fig.~\ref{Fig:E}. These are compared with the $E_g$ bowing from ordered Bi atoms shown in Fig.~\ref{Fig:B}, which gave a slope of $66$~meV/$\%$Bi.

Tab.~\ref{Table:C} and Fig.~\ref{Fig:B}  show that the $E_g$ bowing closest to the $60-90\,$meV/\%Bi range of experimental values found in the literature~\cite{Tixier_APL_82_2003, Huang_JAP_98_2005, Francoeur_APL_82_2003} correspond to the isolated Bi atom and cluster arrangements. Nevertheless, these are not the most favorable total energy states, which may be due to fixing the size of the supercell. These are followed by the bandgap bowing of the $[111]$ chains and of the SQS structures, close to one another but somewhat smaller than the experimental values cited. It is worth noticing that these are our lowest total energy arrangements, and the difference in bowing from the experimental values may be partially due to constraining the size of the supercells. Finally, the $E_g$ bowing for the $[100]$ chains is considerably larger than both the experimental values and the other theoretical values obtained here. 

The spread in the $E_g$ slopes described here suggests that differences in growth techniques or growth parameters can lead to differences in the observed band gap reduction as a function of Bi concentration, which can be caused by the differences in the Bi atom arrangements in the samples. Consequently, measuring the band gap reduction with increasing Bi concentration can potentially help in identifying the types of Bi arrangement distributions in the samples. For example, for growing conditions that fix the lattice constant to that of a GaAs substrate, our results point towards regular Bi distributions or cluster arrangements as the most likely candidates.

For random Bi distributions at high concentrations like that in Fig.~\ref{Fig:A}, the resulting band structure can be understood intuitively as an 'average' of band structures of complexes like those discussed here. Different local configurations would contribute to the effective band structures with weights determined by their total energies. A detailed statistics is beyond the scope of this paper, but the large distributions of $E_g$ is identified  clearly as a factor in the degradation of the Bloch character of the valence band edge seen in Fig.~\ref{Fig:A}.

\subsection{Electronic structure of Bi pairs}

To obtain an insight into the unfolded band structures from Fig.~\ref{Fig:A} and into the distribution of $E_g$ bowing seen in Tab.~\ref{Table:C} and  Fig.~\ref{Fig:E}, we look at the detailed band structure of complexes of two Bi atoms aligned along the $[100]$ axis or along the $[111]$ axis. The former case is relevant for the extraordinary large $E_g$ bowing seen in the previous section for the $[100]$ chains. The latter is relevant for the energy-favorable case of $[111]$ chains, which gives the lowest bowing, albeit one closer to the experimental range. All the calculations from this section are performed with PAW-PP in VASP.

\begin{figure*}
	\includegraphics[scale=.25]{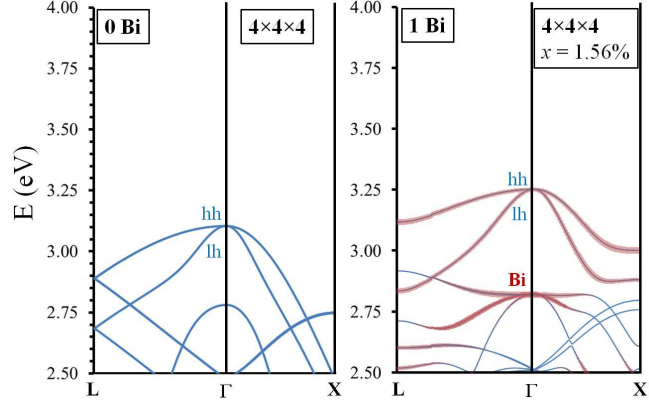}\\
	\caption{The bandstructure of a $4\times 4 \times4$ pristine GaAs supercell compared to the full bandstructure of one Bi atom in a $4\times4\times4$ GaAs host lattice supercell. The widths of the lines are proportional to the contribution of the $p$ orbitals of the Bi atoms. It is seen how the latter contribute significantly to the heavy (hh) and light (lh) branches and how they introduce splittings in these branches at the edges of the Brillouin zone.}\label{Fig:H}
\end{figure*}

\begin{figure*}
	(a)\includegraphics[scale=.40]{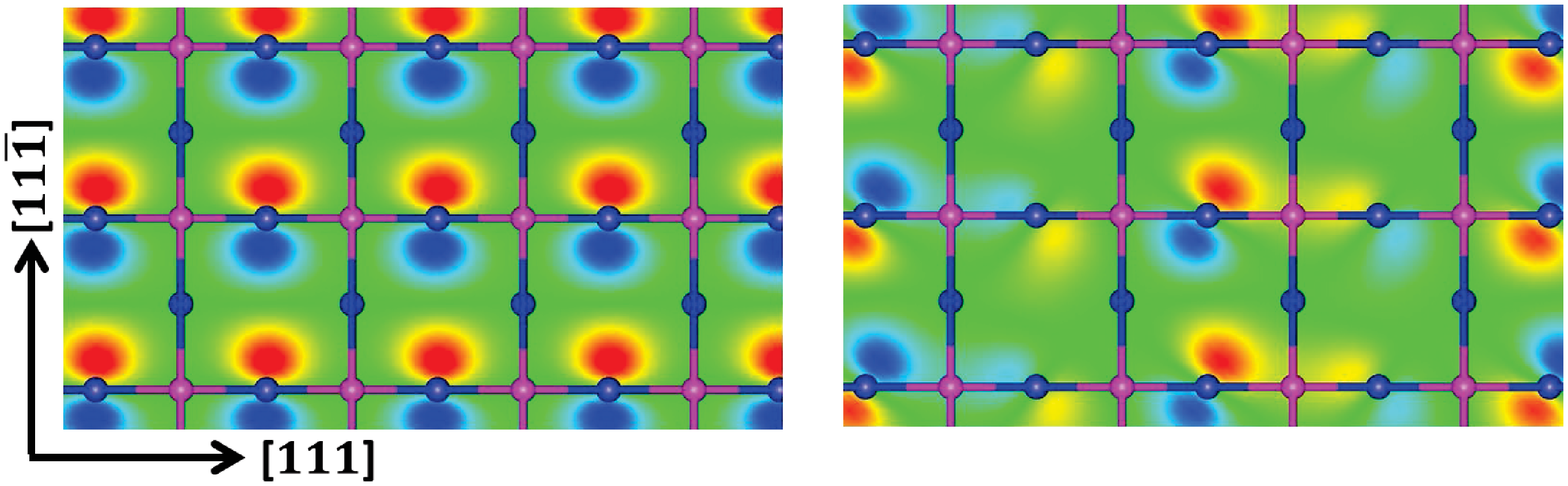}\\
	(b)\includegraphics[scale=.40]{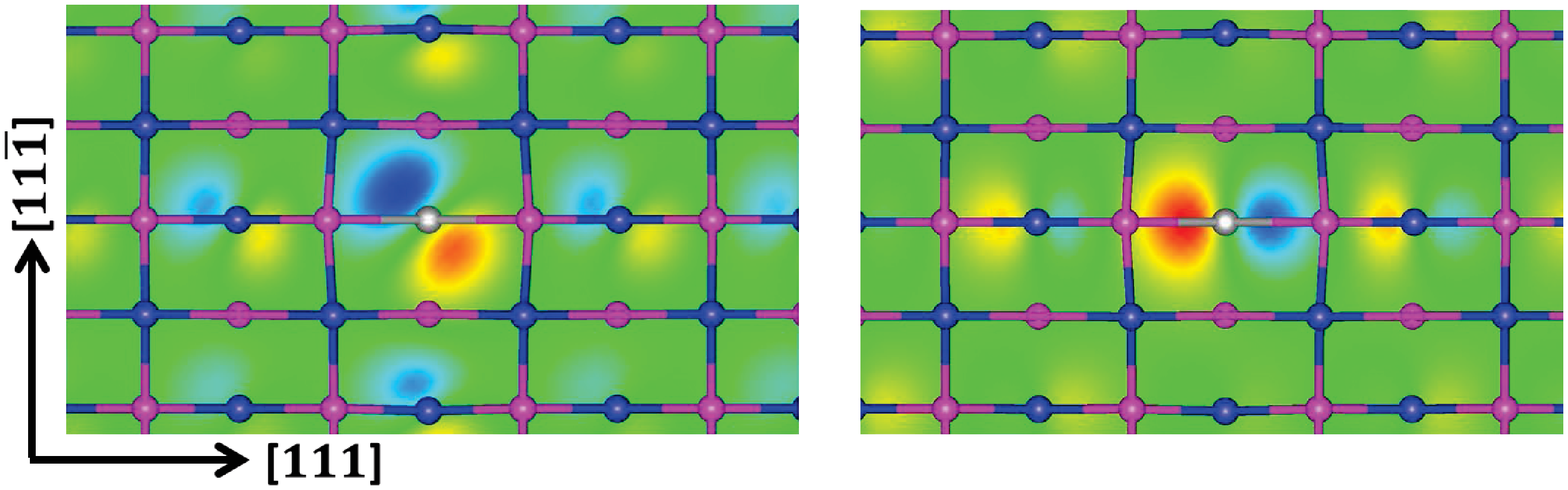}\\
	(c)\includegraphics[scale=.40]{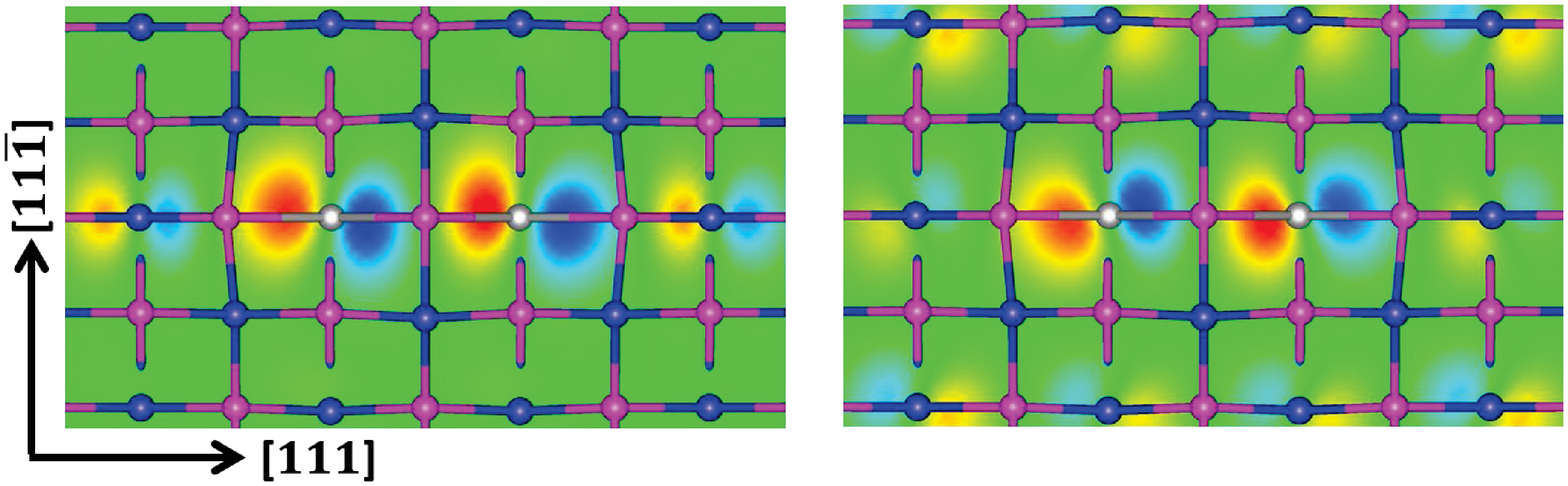}\\
	\caption{Wavefunctions for the $\Gamma$-point Bloch states of: (a) the pristine GaAs, the hh and the so bands; (b) one Bi atom, the hh and the defect level, showing localization and hybridization with the host $p$ orbitals; (c) two Bi atoms in closest proximity along the $[100]$ axis. Only the real part of the 'spin-up' component is shown.}\label{Fig:LNM}
\end{figure*}

First, we obtain the single Bi  band structure shown in Fig.~\ref{Fig:H} for a $128$-atom cell side by side with the folded pristine GaAs band structure. It is seen that the defect level hybridizes strongly with the heavy hole (hh) and light hole (lh) bands and contributes significantly to the valence band edge and to the flat defect level next to the split-off band. There is virtually no hybridization with the split-off band. The splitting of the hh and lh bands gives rise to localized states at the edge of the Brillouin zone, which will contribute to the loss of Bloch character in disordered structures like that in Fig.~\ref{Fig:A}. Fig.~\ref{Fig:LNM}(a,b) shows a comparison between the $\Gamma$-point wavefunctions of pristine GaAs and of a single Bi atom, for the hh state and the defect state. It proves the hybridization of Bi $p$ orbitals with the nearest neighbor As $p$ orbitals and the localization of the state.

\begin{figure*}
	\includegraphics[scale=.3]{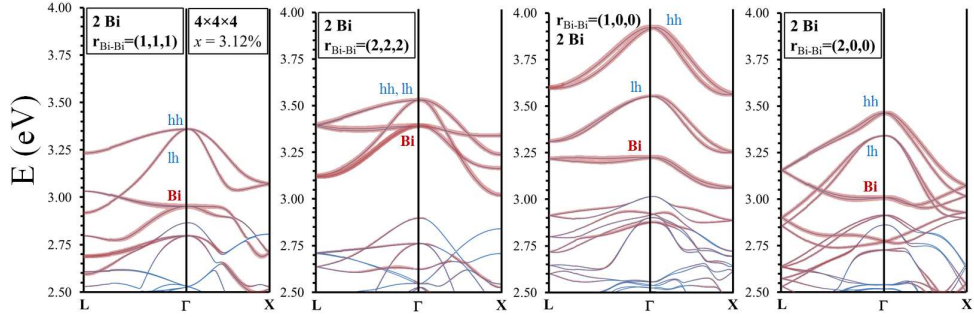}\\
	\caption{The bandstructure of a $4\times 4 \times4$ pristine GaAs supercell compared to the full bandstructure of two Bi atom in a $4\times4\times4$ GaAs host lattice supercell.}\label{Fig:I}
\end{figure*}

Next, we consider the band structure of two Bi atoms in a $128$-atom cell (Fig.~\ref{Fig:I}). There is a clear difference between the alignment along the $[100]$ and the $[111]$ directions: the valence band edge is raised considerably more in the former case. This is due to  different distances between the two Bi atoms: $\approx a_0$ along $[100]$ and $\approx a_0\sqrt{6}$ along $[111]$. The defect levels move away from the so-band, and there is a spin-orbit splitting of hh, lh, and defect bands in the $[100]$ cases due to lower symmetry. Fig.~\ref{Fig:LNM}(c) shows the $\Gamma$-point wavefunctions for the $[100]$ pair, proving the strong overlap between the $p$ orbitals of the two Bi atoms.

\begin{figure*}
	\includegraphics[scale=.25]{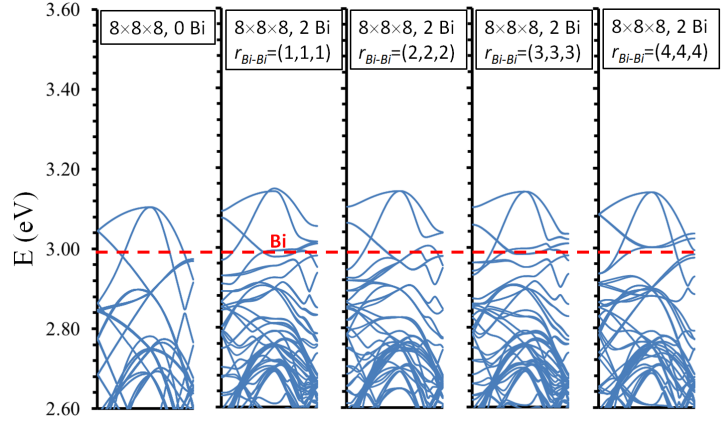}\\
	\caption{The bandstructure of a $8\times 8 \times8$ pristine GaAs supercell compared to the full bandstructure of two Bi atoms aligned along the $[111]$ axis in a $8\times8\times8$ GaAs host lattice supercell. The widths of the lines are proportional to the contribution of the $p$ orbitals of the Bi atoms. The positions of the defect levels and the shifts of the $hh$ and $lh$ states depend strongly on the relative position between the Bi atoms, as discussed in the text.}\label{Fig:J}
\end{figure*}

\begin{figure*}
	\includegraphics[scale=.25]{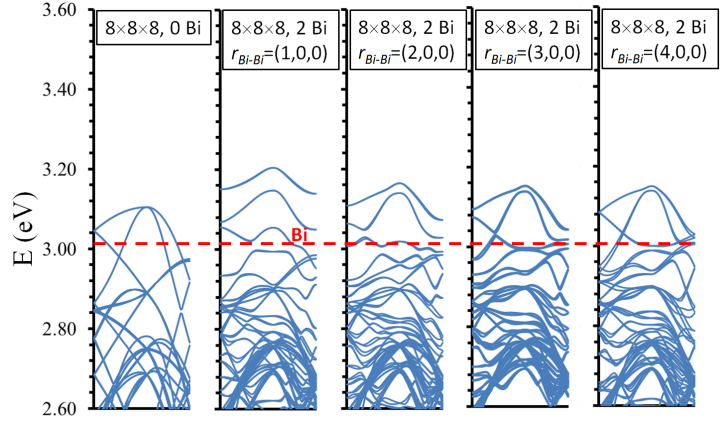}\\
	\caption{The bandstructure of a $8\times 8 \times8$ pristine GaAs supercell compared to the full bandstructure of two Bi atoms aligned along the $[100]$ axis in a $8\times8\times8$ GaAs host lattice supercell.}\label{Fig:K}
\end{figure*}

All configurations shown in (Fig.~\ref{Fig:I}) correspond to the same concentration but give a wide range of $E_g$ bowing values, which shows that the overlaps between $p$ orbitals of Bi is strongly anisotropic. The strong interaction between two neighboring Bi atoms can perturb significantly the valence band even for smaller concentrations, depending on their relative position, as seen in Figs.~\ref{Fig:J} and \ref{Fig:K} for $8\times8\times8$~($x=0.4\%$) calculations with VASP. This size of supercell minimizes the interaction between image Bi atoms. Again, it is seen that a large perturbation occurs in the $[100]$ configuration of the pair. This set of calculations proves that at small concentrations the clustering is minimal, otherwise it would be observed as a larger $E_g$ bowing than that seen in experiments.

\begin{figure*}
	\includegraphics[scale=1.]{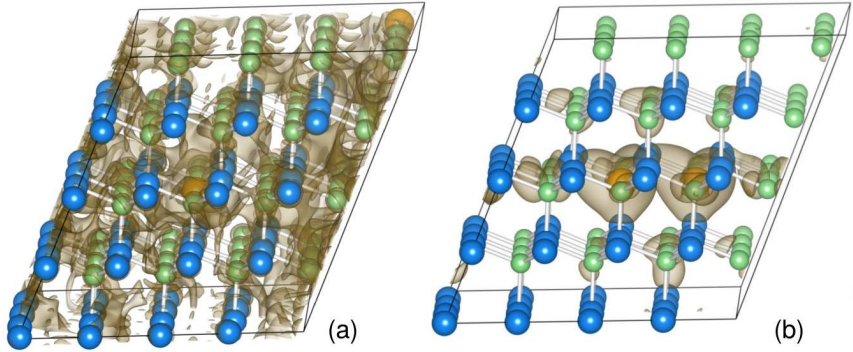}
	\caption{Band decomposed charge density of the heavy hole band for the two atom $[111]$ chain and cluster arrangements from Tab.~\ref{Table:C}. The charge density results from integration over the whole Brillouin zone. Every isovalue is set to 10\%\ of the respective maximum.}\label{Fig:F}
\end{figure*}

The localization and hybridization effects observed in \ref{Fig:LNM}(b,c) are consistent with the real-space interpretation of dilute nitride and bismide band structures \cite{Kent_PRB_64_2001, Virkkala_PRB_88_2013,Virkkala_PRB_88_2013b}. In that interpretation, the main argument is the accumulation of electron charge at single impurity centers. The charge was integrated over the entire defect band. Here we provided a more detailed picture in terms of hybridization and localization of specific wavefunctions for interacting defects. Our argumentation can be complemented with a real-space description of charge accumulation at Bi complexes, which is relevant for the formation of covalent bonds between impurities. This is shown in Fig.~\ref{Fig:F} for the example of the heavy hole band for the two atom $[111]$ chain and pair arrangements from Tab.~\ref{Table:C}. The charge accumulation can be observed in the case of the cluster [Fig.~\ref{Fig:F}(b)] while such an accumulation is absent when the Bi atoms are dispersed [Fig.~\ref{Fig:F}(a)]. 

\section{Conclusions}\label{Sec:Conclusions}

In this work we performed a detailed analysis of three factors influencing the bandgap bowing in dilute $GaAs_{1-x}Bi_x$ alloys: the chemical effect, the strain effect, and the effect of disorder. We found that the strain induced in the lattice by the Bi atoms is responsible for a good part of the bandgap bowing, in particular for large concentrations $x$. To understand the effective band structures at high concentrations, we analyzed the contribution of various cluster configurations to the band bowing. We found that the latter depends strongly on the structure of clusters considered. We provided an understanding of the range of bowing rates observed based on the anisotropic, strongly coordinate-dependent interaction between Bi $p$ atoms. We suggest that the two-scale disorder observed in PL experiments at high Bi concentration can be understood intuitively as coming from an average of valence band perturbations like those seen here, or from an effective band structure with a significant degradation of the 
Bloch character throughout the Brillouin zone. The results from the models analyzed here suggest that some band bowing measurements performed on samples grown in different conditions can be interpreted in terms of special Bi configurations like those studied here.

%
%
\begin{acknowledgments}
The Marburg part of the work was funded by the DFG via the GRK 1782 "Functionalization of Semiconductors"; computing time from HRZ Marburg, CSC Frankfurt, and HLRS Stuttgart is acknowledged. OR would like to acknowledge funding provided by the Natural Sciences and Engineering Research Council of Canada under the Discovery Grant Program RGPIN-2015-04518. The work in USA was supported by the AFOSR, the NLCSTR contributions via a phase-II STTR.
\end{acknowledgments}

%
%
\bibliography{gabiaslit}

\end{document}